\documentclass[twoside]{article}
\usepackage{fleqn,espcrc2}
\usepackage{graphicx}
\usepackage{amsfonts}

\newcommand{\be}{\begin{equation}}
\newcommand{\ee}{\end{equation}}
\newcommand{\bdm}{\begin{displaymath}}
\newcommand{\edm}{\end{displaymath}}
\newcommand{\bea}{\begin{eqnarray}}
\newcommand{\eea}{\end{eqnarray}}
\newcommand{\ba}{\begin{array}}
\newcommand{\ea}{\end{array}}

\newcommand{\pref}[1]{(\ref{#1})}

\title{On Witten's global anomaly for higher SU(2) representations
}
\author{O. B\"ar\address{Center for Theoretical Physics,
        Laboratory for Nuclear Science and Department of 
        Physics,\\
        Massachusetts Institute of Technology (MIT),
        Cambridge, MA 02139, U.S.A.}
        } 
\begin{document}
\begin{abstract}
The spectral flow of the overlap operator is computed numerically along a path connecting two gauge fields which differ by a topologically non-trivial gauge transformation. The calculation  is performed for SU(2) in the 3/2 and 5/2 representation. An even-odd pattern for the spectral flow as predicted by Witten is verified. The results are, however, more complicated than naively expected. 
\vspace{2.8pc}
\end{abstract}

\maketitle

\section{Introduction}
Many years ago Witten gave an argument that some chiral SU(2) gauge theories are mathematically inconsistent \cite{Witten:fp}. He found that  the sign of the fermion determinant cannot be defined satisfying both gauge invariance and smooth gauge field dependence if
\be\label{anomalous_reps}
j\,=\, 2l+\frac{1}{2}\,,\qquad l\,=\,0,1,2\ldots\,,
\ee
where $j$ denotes the highest weight of the SU(2) representation.

His argument is based on the spectral flow of the Dirac operator along the interpolation 
\be\label{cont_path}
A_{\mu}(x,t)\,=\,(1-t)\, A_{\mu}(x) + t\, A_{\mu}^g(x),
\ee
between an arbitrary gauge field $A_{\mu}$ and its gauge transform $A_{\mu}^g\,=\,g(A_{\mu} + \partial_{\mu})g^{-1}$, where the gauge transformation $g$ is an element of the non-trivial class of $\pi_4(SU(2))\,=\, \mathbb{Z}_2$. Taking \pref{cont_path} as the background gauge field for the massless Dirac operator,  $n_j$ eigenvalue pairs $\{\lambda_i(t), \lambda_i^*(t)\}$ cross zero and change places as $t$ is varied between 0 and 1: 
\be\label{sign_flip}
\lambda_{i}\Bigg|_{t=0}\,=\,\lambda^*_{i}\Bigg|_{t=1}\,,\qquad i\,=\,1,\ldots,n_j\,.
\ee
The integer $n_j$ depends on the group representation $j$, and the theory is ill-defined if $n_j$ is an odd integer.

Witten showed, invoking an Atiyah-Singer index theorem \cite{Atiyah:1971} for a certain five-dimensional Dirac operator (the fifth coordinate is essentially the path parameter $t$), that $n_j$ is odd only for the representations in \pref{anomalous_reps}. Besides the fundamental representation with $j=1/2$, the representations with $j=5/2, 9/2,$ etc.~also suffer from Witten's anomaly, while the 3/2 and the remaining half-integer representations, as well as all integer-valued representations, are anomaly free. However, the index theorem does not predict $n_j$ itself. So we can ask the following question: 

What is the value of $n_j$ as a function of $j$ that causes only half of the half-integer representations to be  anomalous? 
\section{Computing $n_j$ numerically}
The integer $n_j$ can be obtained by formulating the path \pref{cont_path} on a discrete space-time lattice and computing numerically the spectral flow along this path for a suitable lattice Dirac operator. Suitable means that the Dirac operator should satisfy the Ginsparg-Wilson relation \cite{Ginsparg:1981bj} in order to preserve exact chiral symmetry even for non-zero lattice spacing. This method has already been employed for the fundamental representation \cite{Bar:1999ka,Neuberger:1998rn}, and $n_{1/2}$ was found to be 1. Here we report about a similar calculation for the $3/2$ and $5/2$ representation and present the results for $n_{3/2}$ and $n_{5/2}$.
\section{Some technical remarks}
We use the overlap operator  \cite{Neuberger:1997fp,Neuberger:1998wv} proposed by Neuberger in our calculation. It is given by
\be\label{def_Neub_op}
\ba{rcl}
aD & = & 1-A(A^{\dagger}A)^{-1/2}\,,\\[0.2ex]
 A & = & 1 - aD_w\,.
\ea
\ee
$D_w$ denotes the usual Wilson--Dirac operator. The inverse square root of $A^{\dagger}A$  
is approximated by a truncated expansion in  Chebyshev  polynomials \cite{Hernandez:1998et}. 
%
\begin{figure}[t]
\begin{center}
\includegraphics*[scale = 0.39, trim=0 35 0 200]{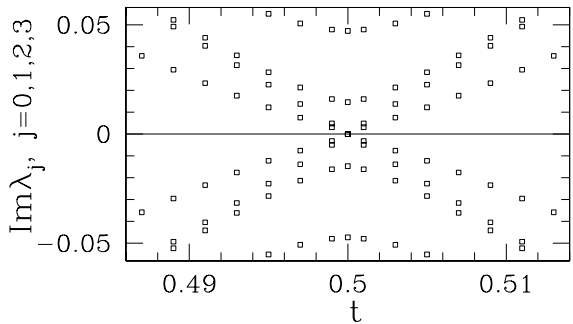}
\vspace{-1cm}
\caption{\label{fig1}\small 
The imaginary parts of the lowest four eigenvalue pairs for the $3/2$ representation.
Two of the pairs cross zero at $t=0.5$, leading to $n_{3/2} = 2$. 
} 
\end{center}
\end{figure}
The Conjugate Gradient algorithm \cite{Bunk94a,Kalkreuter:1995mm} is employed to compute the lowest eigenvalues $|\lambda_i|^2$ of $D^{\dagger}D$. In addition, the CG algorithm gives also the corresponding eigenvectors, which allows the direct computation of the imaginary parts $\mbox{Im}\, \lambda_i$ \cite{Bar:2000qd}. We are therefore able to establish whether the imaginary parts cross zero and change sign or not.

For numerical reasons we choose a constant gauge field as the starting point for the path. This leads to a gap in the spectrum for $t=0$ and 1, which is advantageous in the numerical calculation. The topologically non-trivial gauge transformation $g$ was constructed analytically in the continuum and then restricted to the discrete space--time lattice. A special symmetry property of $g$ leads to the symmetry 
\be\label{symmetry_lambda}
\mbox{Im }\lambda_i (t)\,=\,\pm\mbox{Im }\lambda_i(1-t)\,.
\ee
for the eigenvalues as a  function of $t$. Consequently, the imaginary parts of those $n_j$ pairs satisfying \pref{sign_flip} cross zero at $t=0.5$. In order to determine $n_j$ it is therefore sufficient to compute the spectral flow for a small neighborhood around $t=0.5$.
\begin{figure}[t]
\begin{center}
\includegraphics*[scale = 0.385, trim=0 35 0 200]{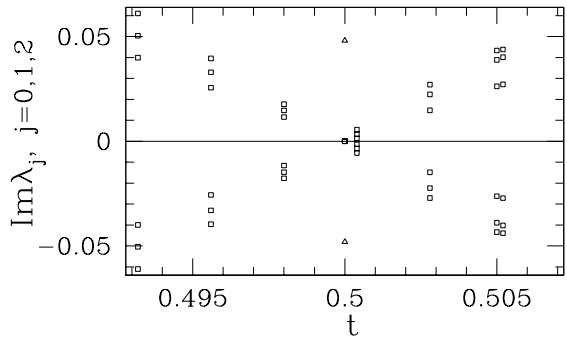}
\vspace{-1cm}
\caption{\label{fig2}\small 
The imaginary parts for the $5/2$ representation. Three eigenvalue pairs cross zero. The additional triangular data point at $t=0.5$ corresponds to the fourth eigenvalue. It is unequal to zero and therefore the fourth pair does not  cross zero.
} 
\end{center}
\end{figure}

\section{Results} 
Figs.~\ref{fig1} and \ref{fig2} show the numerical results. We find $n_{3/2}=2$ and $n_{5/2}=3$, which is in agreement with Witten's even-odd prediction based on the index theorem. 

The results for the lowest three half-integer representations  can be summarized by the simple relation
\be\label{my_conjecture}
n_j\,=\, j + \frac{1}{2}\,.
\ee
Whether this relation holds true for all representations is not known. Of course, a rigorous proof of \pref{my_conjecture} for all $j$ is beyond the scope of a numerical calculation. However, \pref{my_conjecture} is the simplest relation one can imagine that gives rise to the anomaly pattern \pref{anomalous_reps} in terms of the spectral flow. 

\begin{figure}[t]
\begin{center}
\includegraphics*[scale = 0.38, trim=0 35 0 200]{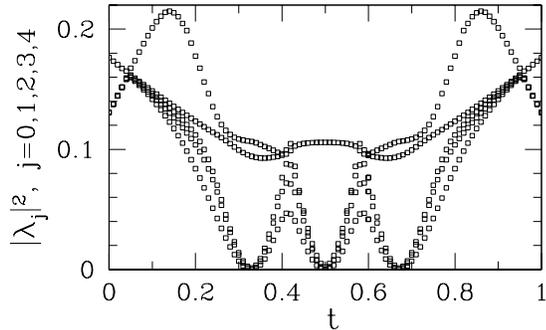}
\vspace{-1cm}
\caption{\label{fig3}\small 
The lowest five eigenvalues of $D^{\dagger}D$ for the 3/2 representation. The symmetry $|\lambda_i|^2(t)=|\lambda_i|^2(1-t)$ is a consequence of property \pref{symmetry_lambda}. For $t\approx0.33\,,0.5$ and $0.67$ three eigenvalues come close to zero.
} 
\end{center}
\end{figure}

\begin{figure}[t]
\begin{center}
\includegraphics*[scale = 0.39, trim=0 36 0 200]{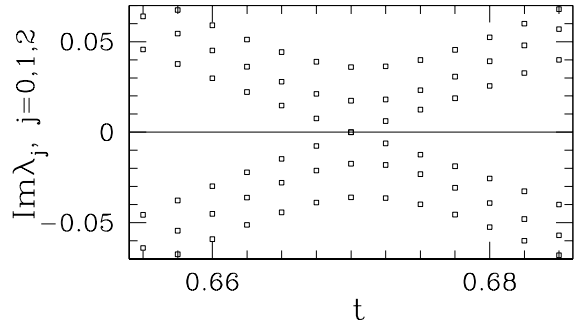}
\vspace{-1.1cm}
\caption{\label{fig4} \small 
The imaginary parts of the lowest three eigenvalue pairs for $j=3/2$ around $t=0.67$.
One of the pairs crosses zero at $t\approx 0.67$. Because of the symmetry property \pref{symmetry_lambda} this implies an additional crossing at $t\approx 0.33$.
} 
\end{center}
\end{figure}

In addition to our results for $n_{3/2}$ and $n_{5/2}$  we find some unexpected behavior for the spectral flow. In order for an eigenvalue pair to change places, it must cross zero at least once along the path. The total number of zero crossings, however, is larger than necessary. For the 3/2 representation we find 4 zero crossings, one eigenvalue pair crosses zero 3 times, the second one once (see figs.~\ref{fig3} and \ref{fig4}). 

Along the same path with SU(2) in the 5/2 representation we find 9 zero crossings \cite{obaer}. The results suggest that one eigenvalue pair crosses 5 times, another one 3 times and a third one once, even though our numerical approach could not establish this beyond any doubt.

For the 3/2 representation the computation was repeated with 10 randomly generated gauge transformations in the non-trivial class of $\pi_4(SU(2))$. We always found 4 zero crossings, never the naively expected 2. Unfortunately, the same check could not be performed for the 5/2 representation for lack of computer resources.

The results for the total number of zero crossings $k_j$ along the path \pref{cont_path} can be summarized by the relation:
\be\label{conjecture_2}
k_j\,=\,\left(j+\frac{1}{2}\right)^2\,=\, n_j^2\,.
\ee
Even though it is tempting to propose that this relation is valid in general, one should be careful.
In contrast to $n_j$ it is not yet clear whether the number of zero crossings is really a path independent quantity. For further discussion see \cite{obaer}.

This work is supported in part by funds provided by the U.S Department
of Energy (D.O.E.) under cooperative agreement DE-FC02-94ER40818.



\begin{thebibliography}{20}
\bibitem{Witten:fp}
E.~Witten,
Phys.\ Lett.\ B {\bf 117}, 324 (1982).

\bibitem{Atiyah:1971}
M.F.~Atiyah and I.M.~Singer,
Ann.~of Math.\ {\bf 93} (1971) 139.


\bibitem{Ginsparg:1981bj}
P.~H.~Ginsparg and K.~G.~Wilson,
Phys.\ Rev.\ D {\bf 25}, 2649 (1982).

\bibitem{Bar:1999ka}
O.~B\"ar and I.~Campos,
Nucl.\ Phys.\ Proc.\ Suppl.\  {\bf 83}, 594 (2000).

\bibitem{Neuberger:1998rn}
H.~Neuberger,
Phys.\ Lett.\ B {\bf 437}, 117 (1998).

\bibitem{Neuberger:1997fp}
H.~Neuberger,
Phys.\ Lett.\ B {\bf 417}, 141 (1998).

\bibitem{Neuberger:1998wv}
H.~Neuberger,
Phys.\ Lett.\ B {\bf 427}, 353 (1998).

\bibitem{Hernandez:1998et}
P.~Hernandez, K.~Jansen and M.~Luscher,
Nucl.\ Phys.\ B {\bf 552}, 363 (1999).

\bibitem{Bunk94a}
B.~Bunk, K.~Jansen,  M.~L\"uscher and  H.~Simma,
\newblock ALPHA collaboration internal report (unpublished), 1994.

\bibitem{Kalkreuter:1995mm}
T.~Kalkreuter and H.~Simma,
Comput.\ Phys.\ Commun.\  {\bf 93}, 33 (1996).

\bibitem{Bar:2000qd}
\newblock O.~B\"ar,
DESY-THESIS-2000-022.

\bibitem{obaer}
\newblock O.~B\"ar, in preparation.

\end{thebibliography}
\end{document}